\documentclass[12pt]{JHEP3}

\usepackage{amsmath}
\usepackage{amsfonts}
\usepackage{eufrak}
\usepackage{amssymb,bbold}
\usepackage{epsfig}
\usepackage{bm}

\def\be{\begin{equation}}
\def\ee{\end{equation}}

\def\a{\alpha}
\def\b{\beta}
\def\n{\nabla}
\def\t{\tau}

\def\v{\nu}
\def\l{\lambda}
\def\m{\mu}
\def\s{\sigma}

\def\o{\omega}

\def\pa{\partial}

\def\O{\Omega}
\def\Ob{\overline{\Omega}}
\def\e{\epsilon}
\def\G{\Gamma}

\def\sq{\sqrt{2}}
\def\d{\delta}
\def\mr{\mathfrak{R}}

\def\W{\mathcal{W}}

\def\bfI{\mbox{\boldmath $I$}}
\def\6{\partial}
\newcommand{\eref}[1]{(\ref{#1})}

\newcommand{\ip}{\raise1pt\hbox{\large$\lrcorner$}\,}
\newcommand{\bea}{\begin{eqnarray}}
\newcommand{\eea}{\end{eqnarray}}
\newcommand{\nn}{\nonumber \\}

\title{Timelike Killing spinors in seven dimensions}

\author{Marco Cariglia \thanks{M.Cariglia@damtp.cam.ac.uk} and
  Ois\'{\i}n A. P. Mac
  Conamhna\thanks{O.A.P.MacConamhna@damtp.cam.ac.uk} \\ DAMTP \\ Centre
  for Mathematical Sciences \\ University of Cambridge \\ Wilberforce
  Road, Cambridge CB3 0WA, UK.}

\abstract{We employ the G-structure formalism to study supersymmetric
  solutions of minimal and $SU(2)$ gauged supergravities in seven
  dimensions admitting Killing spinors with associated
  timelike Killing vector. The most general such Killing spinor
  defines an $SU(3)$ structure. We deduce necessary and sufficient
  conditions for the existence of a timelike Killing spinor on the
  bosonic fields of the theories, and find that such configurations
  generically preserve one out of sixteen supersymmetries. Using our
  general supersymmetric ansatz we obtain numerous new
  solutions, including squashed or deformed AdS solutions of the
  gauged theory, and a large class of G\"{o}del-like solutions with
  closed timelike curves.}
 
\preprint{DAMTP-2004-74}

\begin{document}

\section{Introduction}

In recent years, gauged supergravities in various dimensions have been
recognised as powerful technical tools for the construction of
interesting supersymmetric backgrounds of string/ M theory. A
particularly fruitful application has been to the generalised AdS/CFT
correspondence. Following the celebrated work of Maldacena and Nunez,
\cite{carlos}, \cite{ccarlos},
an extensive literature has developed on branes wrapping various
supersymmetric cycles, and their associated field theories; a review
and examples of applications 
of the supergravity aspects are to be found in \cite{review}-\cite{F4}. On the
gravity side, the near-horizon geometries are most easily constructed
in some lower dimensional gauged supergravity, before lifting to
$d=10$ or 11. In view of this and other applications, it is clearly
important to have some systematic understanding of supersymmetric
solutions in gauged supergravities.

More generally, the ability to map out all supersymmetric backgrounds
in any conceivable (supergravity) limit of string/ M theory is 
of much value. In fact, the framework which allows one to achieve
precisely this has been identified, and is provided by the notion of a
G-structure. Already, much work has been done in various contexts,
\cite{tod}-\cite{klemm}, and
a classification of all supersymmetric solutions of $d=11$
supergravity has been given \cite{Gauntlett_Pakis}, \cite{gaunt3}. 

Geometrically, G-structures provide information about 
supersymmetric field configurations. Typically one finds that fluxes
in a given theory are tightly constrained (and sometimes entirely
fixed) by the torsion classes of the structure. However, it must be
stated that the approach is not 
without its drawbacks. Generally speaking, the classification becomes
progressively more implicit as the dimensionality of spacetime and/ or
the number of supercharges increases. The point is that while the
existence of a G-structure implies and is implied by the vanishing of
the supersymmetry variations of the fermions in the theory, including
the Killing spinor equation, it is not equivalent to having a solution
of the field equations and Bianchi identities. When a Killing
spinor exists, generically some but not all of these are satisfied
identically. The remainder must be imposed as additional constraints
on the bosonic fields, over and above those implied by the
G-structure. Typically, they take the form of differential constraints
on the torsion classes of the structure, and these are often
hard to solve. Nevertheless, the generality of the technique, and its
constructive nature in providing a clear set of prescriptions for constructing
explicit solutions, makes it very powerful.

In this paper we will apply the G-structure formalism to minimal and
$SU(2)$ gauged supergravity in seven dimensions. We will focus on
configurations admitting a timelike Killing spinor - that is, a
Killing spinor whose associated Killing vector is timelike. The null
case will be the subject of a future work. We find that a timelike
Killing spinor is equivalent to an $SU(3)$ structure, and determine the
necessary and sufficient conditions on the bosonic fields for its
existence. 

One of the more surprising and unsettling features of
supergravities revealed by G-structures is the extent to which the
class of (particularly timelike) supersymmetric solutions is infested
by spacetimes with closed timelike curves. We find many new examples
here, in the gauged theory as bundles over negative scalar curvature
K\"{a}hler threefolds, which lift to $d=10$, 11. We show how various
$AdS_{7,5,3}$ solutions of the theory, for which the closed timelike
curves may be eliminated, arise as special cases of this much broader
class of solutions, and also how the AdS factors may be squashed or
deformed by the addition of suitable fluxes.

The plan of the paper is as follows. In section two we
describe the 
theories we study. In section three we give a summary of the necessary
and sufficient conditions for a bosonic field configuration in the
theories to admit a timelike Killing spinor. The derivation of these
conditions is quite technical, and is relegated to the appendices. In
sections four and five we obtain explicit supersymmetric solutions of the
minimal and gauged theories respectively. Section six concludes. In
appendix A we give our conventions, and miscellaneous
useful material. In appendix B we compute the various bilinears that
may be constructed from a Killing spinor, and show how they define an
$SU(3)$ structure in seven dimensions. In appendix C we derive the 
necessary conditions for a bosonic configuration to admit a timelike
Killing spinor, and show that these are also sufficient. In appendix
D we discuss the intrinsic torsion and contorsion of an $SU(3)$
structure in six Riemannian dimensions. In appendix F we give the
integrability conditions for the theories. 

\section{The theory and supersymmetry variations \label{sec:theory}}
The lagrangian for minimal seven dimensional ungauged supergravity was first
written down in \cite{Townsend:1983kk}.  The $SU(2)$ gauged version was also
written down in \cite{Townsend:1983kk} but with numerical typos, which
were corrected in \cite{Mezincescu:ta}.  In \cite{Townsend:1983kk}
Euclidean signature is used. 
\subsection{The minimal theory}
The
bosonic lagrangian density for the minimal theory in our conventions is
\bea
\mathcal{L} & = & \frac{1}{2} e R - \frac{1}{24 } e
(G_{\mu\nu\rho\tau})^2  - \frac{1}{2} e  F_{\mu\nu}\,_{\;\;b}^a
F^{\mu\nu }\,^b_{\;\;a} - \frac{5}{2}e (\6_\mu \phi)^2  \nn
&& - \frac{1}{24}ee^{-\phi}
G_{\mu\nu\rho\tau} F_{\kappa \lambda }\,^a_{\;\;b} A_{\chi }\,^b_{\;\;a}
\e^{\mu\nu\rho\tau\kappa \lambda\chi} \, . 
\eea
Compared to \cite{Townsend:1983kk} we use the same conventions for the
Riemann tensor but Hawking and Ellis conventions for the Ricci tensor
and scalar.  We have also rescaled $\phi\rightarrow {\sqrt 5}\, \phi$
and the forms by $F\rightarrow {\sqrt 2}\,e^{\phi} F$, $G\rightarrow
{\sqrt 2}\,e^{-2\phi} G$, $A_{(1),(3)}\rightarrow\sq A_{(1),(3)}$. 
The supersymmetry
variations of the fermions are given by
\bea
    \delta \lambda^a & = & \frac{\sqrt{5}}{2} \G^\mu D_\mu \phi \, \e^a +
    \frac{i}{2\sqrt{5}}  \G^{\mu\nu} F_{\mu\nu }\,^a_{\;\;b} \, \e^b +
    \frac{1}{24\sqrt{5}} 
    \G^{\mu\nu\rho\tau} G_{\mu\nu\rho\tau} \, \e^a ,  \\ 
    \delta \psi_{\mu }^a & = & D_\mu \, \e^a - \frac{i}{10} 
    (\G_\mu\,^{\nu\rho}-8 \delta^\nu_\mu \G^\rho) F_{\nu\rho }\,^a_{\;\;b} \,
    \e^b + \frac{1}{80} 
    (\G_\mu\,^{\alpha\beta\gamma\delta}-\frac{8}{3}
    \delta^\alpha_\mu \G^{\beta\gamma\delta})
    G_{\alpha\beta\gamma\delta}\, \e^a \, , \label{eq:Killing} 
\eea
and the parameter $\e^a$ is a symplectic-Majorana spinor, whose
properties are summarized in appendix A.  

Let us introduce the following notation. Let $A_p$, $B_q$ be $p$-
and $q$-forms respectively. Then  
\be
A\lrcorner
B_{a_1...a_{q-p}}=\frac{1}{p!}A^{b_1...b_p}B_{b_1...b_pa_1...a_{q-p}}.
\end{equation}
The equations of motion and Bianchi identities are
\begin{eqnarray}
   d(e^{-2\phi}G) &=& 0, \\ 
   d(e^{\phi}F^A) &=& 0,\\  
   P &=& 5\n^2\phi-4G\lrcorner G+F^A\lrcorner F^A=0, \label{eq:P} \\ 
   Q &=& \star(e^{-2\phi}d\star(e^{2\phi}G)-\frac{1}{2}F^A\wedge
   F^A)=0, \label{eq:Q} \\ 
   R^A &=& \star(e^{\phi}d\star(e^{-\phi}F^A)-2F^A\wedge G)=0,
   \label{eq:R} \\
   E_{\m\v}&=&R_{\m\v}-\frac{1}{3}\Big(G_{\m\a\b\gamma}G_{\v}^{\;\;\;\a\b\gamma}-
   \frac{1}{10}g_{\m\v}G_{\a\b\gamma\d}G^{\a\b\gamma\delta}\Big)-5\pa_{\m}\phi
   \pa_{\v}\phi \nonumber\\
   &-& \left(
   F^{A}_{\m\a}F^{A\a}_{\v}-\frac{1}{10}g_{\m\v}F^A_{\a\b}F^{A\a\b}
   \right) \nonumber = 0.
\eea
The minimal theory arises as a truncation of type I supergravity
compactified on a $T^3$, or $d=11$ supergravity compactified on K3
\cite{Townsend:duality}.

\subsection{The gauged theory}
To gauge the theory, one covariantises with respect to $SU(2)$,
\be
e^{\phi}F^A=dA^A+\frac{g}{2}\e^{ABC}A^B\wedge A^C,
\end{equation}
and adds the following terms to the bosonic Lagrangian density:
\be
\d\mathcal{L}=e(-V(\phi)+8he^{-2\phi}\star(G\wedge A_{(3)})).
\end{equation}
The potential is given by
\be
V(\phi)=-60m^2+10(m^{\prime})^2,
\end{equation}
where $m$ is a function of the single scalar field $\phi$,
\be
m=-\frac{2}{5}he^{-4\phi}-\frac{1}{10}ge^{\phi},
\end{equation}
with $g$ the gauge coupling (we have rescaled the coupling in
\cite{Townsend:1983kk} by $g\rightarrow g/\sq$)
and $h$ the (constant) topological
mass. One modifies the supersymmetry transformations of the fermions
by adding the terms
\bea
\d\psi_{\m}^a|_{gauge}  &=&m\G_{\m}\e^a-igA_{\m\;b}^{\;a}\e^b,\\
\d\lambda^a|_{gauge} &=&-\sqrt{5}m^{\prime}\e^a.
\eea
In addition to $SU(2)$ covariantisation, the field equations are
modified by the addition of the following terms: 
\bea
 P|_{gauge}&=&-V^{\prime},\\ Q|_{gauge}&=&8he^{-4\phi}\star G,\\
E_{\m\v}|_{gauge}&=&-\frac{2}{5}g_{\m\v}V
\eea
When $h=0$, the $SU(2)$ theory lifts on an $S^3$ to the NS sector in
$d=10$ \cite{chamseddine}. When $hg>0$, it lifts on an $S^4$ to $d=11$
\cite{pope}. When $h\neq 0$, there is a subtlety in imposing the
four-form field equation. The reason is that the 3-form $A_{(3)}$,
which is massive, would have twenty on-shell degrees of freedom if it
satisfied an ordinary second order field equation. However the 3-form
in the 7d supergravity multiplet should have only ten on-shell degrees
of freedom. This is achieved by imposing the odd-dimensional
selfduality equation \cite{Pilch}:
\be\label{selfd}
e^{2\phi}\star G-\frac{1}{2}(e^{\phi}F^A\wedge
A^A-\frac{g}{6}\e^{ABC}A^A\wedge A^B\wedge A^C)+8hA_{(3)}=0.
\end{equation}
Note that the exterior derivative of this equation is just $\star
Q$. Imposing the Bianchi identity and $Q=0$ fixes $A_{(3)}$ up to an
arbitrary closed three form. The closed three form is then determined
by demanding that $A_{(3)}$ satisfies (\ref{selfd}). In the examples
given below, we will explicitly impose the Bianchi identity and $Q=0$,
but leave the determination of the closed three form in $A_{(3)}$
implicit.

\section{Necessary and sufficient conditions for supersymmetry}
We present here a set of necessary and sufficient conditions
corresponding to having at least one Killing
spinor. The reader interested in the derivation of these conditions
may consult appendices B-D. We start with
the minimal theory and then at the end of the section present the
results for the gauged case.  

The general metric is given by 
\be\label{gmetric}
ds^2=-H^2(dt+\o)^2+g_{ij}dx^idx^j,
\end{equation}
where there is no $t$ dependence in $H$, $\omega$,
$g_{ij}$. $\frac{\pa}{\pa t}$ is Killing with associated one-form 
\be
V=-H^2(dt+\o).
\end{equation}
There is a natural decomposition over a $6$--dimensional Riemannian manifold with
metric $g_{ij}dx^i dx^j$ which we will refer to as the base
henceforth. The base admits an $SU(3)$ structure, according to appendix
B, with an almost complex structure $J$ and a holomorphic 3--form
$\Omega$. We may choose the basis on the base such that these take the
canonical form
\bea
J&=&e^{12}+e^{34}+e^{56},\\\O&=&(e^1+ie^2)\wedge(e^3+ie^4)\wedge(e^5+ie^6).
\eea
Redefine
the two-form field strengths according to
\bea
\tilde{F}^a_{\;\;\;b}&=&(S^{-1}FS)^a_{\;\;\;b},\\K&=&\tilde{F}^3,\\L&=&\tilde{F}^1-i\tilde{F}^2,
\eea
where $S$ is the $SU(2)$ matrix given by equations
\eref{eq:spinor_decomposition3_pre}, \eref{eq:spinor_decomposition3}
and the subsequent discussion. Also
define
\be
C_{\m }\,^a_{\;\;\;b}=i(S^{-1})^a_{\;\;\;c}\,\pa_{\m}S^{c}_{\;\;\;b}.
\end{equation}
Then the
most general expression for the matter fields of 
the minimal theory given the existence of a timelike Killing spinor is as
follows:
\bea
   \phi &=& \phi(x) \label{eq:ns_phi}  \\
   K &=& -\frac{1}{H}V\wedge d(\log(He^{\phi}))+
   \frac{1}{4}\Big(i(C^1-iC^2)\lrcorner\O+c.c.\Big)-Hd\o+\tilde{K}^{(1,1)_0}, \\ 
   L &=& -\frac{i}{H}V\wedge(C^1-iC^2) \hspace{-.01cm}  -
   \hspace{-.01cm}  \frac{1}{4}(d\log H+2\W_4)\lrcorner
\overline{\O}-2i\overline{\W}_1J-i\overline{\W}_2+L^{(2,0)},\\ 
  G &=& -\frac{e^{2\phi}}{2H^2}V\wedge
d(He^{-2\phi}J)\nonumber\\&+&\star_6\Big(-\frac{H}{2}d\o+\tilde{K}^{(1,1)_0}
+\frac{1}{4}\Big(i(C^1-iC^2)\lrcorner\O+c.c.\Big)\Big). \label{eq:ns_G} 
\eea
The forms are almost entirely fixed in terms of the metric, the
dilaton and the $d=6$ structure. The only unconstrained components are
$\tilde{K}^{(1,1)_0}$ and $L^{(2,0)}$, which are arbitrary. The
torsion modules, whose definition is given in appendix
\ref{app:nec_and_suff}, are given by 
\begin{eqnarray}
  \W_1 &=& - \frac{i}{6} \overline{L}\lrcorner \, J , \label{eq:W1}  \\ 
  \W_2 &=& -i \overline{L}^{(1,1)_0} , \\ 
  \W_{3 ijk} &=& 2 G^{(2,1)_0}_{0ijk} , \\ 
  \W_{4 i} &=& \frac{1}{2} G_{0ilm} J^{lm} - \frac{e^{2\Phi}}{H}\pa_i
  \left( He^{-2\Phi} \right) , \\ 
  \W_{5 i} &=& -H^{-1}\pa_i H +\pa_i \Phi
  +J_i^{\;\;j}C^3_j . \label{eq:W5}
\end{eqnarray}   
These conditions give the general supersymmetric ansatz for the field
equations. Of course since we have a Killing spinor some of the field
equations will be identically satisfied. As we show in the appendix
\ref{app:integrability},
it is sufficient to impose the Bianchi identities for the forms, and
the four form and dilaton field equations. The integrability
conditions then guarantee that the remaining field equations are
satisfied.

Now we discuss the gauged theory. The metric is the same, and in 
appendix C we show that there exists a gauge such that the matter fields
are given by 
\bea
   \phi &=& \phi(x)  \label{eq:ns_phi_g} \\
   K &=& -\frac{1}{H}V\wedge d(\log(He^{\phi}))+ 
   \frac{1}{4}\Big(ig(A^1-iA^2)\lrcorner\O+c.c.\Big)-Hd\o+\tilde{K}^{(1,1)_0} 
   \nonumber\\ 
   &+& \frac{1}{3}(8he^{-4\phi}+ge^{\phi})J, \\ 
   L &=& -\frac{ig}{H}V\wedge(A^1\hspace{-.1cm}
   -iA^2)\hspace{-.05cm}-\frac{1}{4} (d\log
   H+2\W_4)\lrcorner
   \overline{\O}-2i\overline{\W}_1J-i\overline{\W}_2+L^{(2,0)}, \\ 
  G &=& -\frac{e^{2\phi}}{2H^2}V\wedge d(He^{-2\phi}J)\nonumber\\ 
  +  &\hspace{-.15cm} \star_6& \hspace{-.15cm}
  \Big( \hspace{-.15cm}  -\frac{H}{2}d\o+\tilde{K}^{(1,1)_0} 
+\frac{1}{4}\Big(ig(A^1-iA^2)\lrcorner\O+c.c.\Big) \hspace{-.1cm} -
\hspace{-.1cm} \frac{1}{3}(4he^{-4\phi}-ge^{\phi})J\Big).  
\eea      
Again there are two arbitrary forms, $\tilde{K}^{(1,1)_0}$ and
$L^{(2,0)}$. The first four torsion modules of the six dimensional
$SU(3)$ structure are the same as for the
ungauged case while the fifth is given by 
\be
\W_{5i}=-H^{-1}\pa_iH+\pa_i\phi+gJ_i^{\;\;j}A^3_j. \label{eq:W5_g} 
\end{equation}
As before, in order to ensure a solution of the field equations, it is
sufficient to impose the Bianchi identities, the four form field
equation and the dilaton field equation. Now we turn to an
investigation of the supersymmetric 
solutions of the theories.

\section{Supersymmetric solutions of the minimal theory}
In this section we will employ our general supersymmetric ansatz for
the minimal theory to
determine some new solutions. The stategy is to make a choice for the
base which satisfies the required constraints on the torsion, and to
use the field equations to determine the field components  which are not fixed by
the (six dimensional) torsion. The general problem is still
complicated, so we will restrict attention to some specific types of
base. 

\subsection{Calabi-Yau base}
The simplest choice for the base is to take it to be
Calabi-Yau. Rather trivially, we see that all the vacuum solutions of
the theory are of the form $\mathbb{R}\times CY_6$. More
interestingly, we may obtain G\"{o}del solutions by looking for
solutions of the form $L=\tilde{K}^{(1,1)_0}=\phi=0$,
$H=1$\footnote{When $H$ and $\phi$ are constant, we may without any
  essential loss of generality set $H=1$, $\phi=0$ in both the minimal
  and gauged theories by performing
  suitable rescalings.}. The Bianchi identity for $G$ then implies
that $d\o$ is coclosed, and it is a simple matter to check that the
dilaton and four form equations of motion are satisfied. Explicitly,
one might choose a flat base with metric $\d_{ij}dx^idx^j$, and $\o$
of the form
\be
\o=\a\sum_{n=0}^{2}(x^{2n+1}dx^{2n+2}-x^{2n+2}dx^{2n+1}),
\end{equation}
and the fluxes are
\bea
K&=&-2\a J,\\G&=&-\frac{1}{2}J\wedge J.
\eea
There are closed timelike curves for
$(x^{2n+1})^2+(x^{2n+2})^2>\a^{-2}$. More generally, we could have
included $(2,0)+(0,2)$ and $(1,1)_0$ parts in $d\o$.

\subsection{Semi-K\"{a}hler base}
A semi-K\"{a}hler base is one for which the only nonvanishing torsion
module is $\W_3$. We will seek solutions induced by such a base with
$H=1$, $\phi=0$. The forms then reduce to
\bea
G&=&-\frac{1}{2}V\wedge\W_3+\star_6(-\frac{1}{2}d\o+\tilde{K}^{(1,1)_0}),\\K
&=&-d\o+\tilde{K}^{(1,1)_0},\\L&=&L^{(2,0)}.
\eea
An explicit example of a semi-K\"{a}hler threefold is the three
dimensional complex Heisenberg group. It admits a left-invariant
metric with structure equations
\bea
de^a&=&0,\;\;a=1,..,4,\\de^5&=&e^{13}-e^{24},\\de^{6}&=&e^{14}+e^{23},
\eea
so that
\bea
dJ&=&e^{136}-e^{246}-e^{145}-e^{235}=\W_3,\\d\O&=&0.
\eea
Now, defining
\be
\{x^i,x^j\}=\frac{1}{2}(x^idx^j-x^jdx^i),
\end{equation}
we may introduce coordinates according to
\bea
e^a&=&dx^a,\;\;a=1,..,4,\\e^5&=&du+\{x^1,x^3\}-\{x^2,x^4\},\\e^6&=&dv+\{x^1,x^4\}+\{x^2,x^3\}.
\eea
The Bianchi identities for the two forms imply that $\tilde{K}^{(1,1)_0}$ and
$L^{(2,0)}$ are closed. We will choose
\bea
\tilde{K}^{(1,1)_0}&=&\a(e^{13}+e^{24})+\b(e^{14}-e^{23}),\\L^{(2,0)}&=&\gamma[e^{13}-e^{24}+i(e^{14}+e^{23})],
\eea
for constant $\a$, $\b$, $\gamma$, and $\gamma$ complex. The four form Bianchi identity may
then be solved by taking
\bea
\tilde{K}&=&d\o^{(1,1)_0},\\d\o^{(0,0)}&=&0,\\d(d\o^{(2,0)+(0,2)})&=&0,
\eea
and so we may choose
\be
d\o^{(2,0)+(0,2)}=\d (e^{13}-e^{24})+\e(e^{14}+e^{23}),
\end{equation}
for constant $\d$, $\e$. Hence $\o$ is given by
\be
\o=(\a+\d)\{x^1,x^3\}+(\a-\d)\{x^2,x^4\}+(\b+\e)\{x^1,x^4\}+(-\b+\e)\{x^2,x^3\}.\end{equation}
The four form field equation imposes
\be
\a^2+\b^2=2(|\gamma|^2+1),
\end{equation}
and it may be verified that this is equivalent to the dilaton field
equation, so there are no further constraints.

\subsection{Hermitian base} 
Manifolds in the class $\mathcal{W}_3\oplus\mathcal{W}_4$ are called
Hermitian, as they have vanishing Nijenhuis tensor. Again we
consider solutions for which $H=1$, $\phi= 0$, $C =0$. Then we have 
\bea
   K &=& -d\o +\tilde{K}^{(1,1)_0} , \\
   L &=& -\frac{1}{2} \W_4 \, \lrcorner \, \Ob + L^{(2,0)}, \\
   G &=& \frac{1}{2} (dt + \o) \wedge dJ + \star_6 \left( -\frac{1}{2}
   d\o + \tilde{K}^{(1,1)_0} \right) . 
\eea 
As an example we take the fibration of a two--dimensional flat space
(a torus or a plane) over another four--dimensional flat space: 
\bea 
   d e^5 &=& e^{12} , \\
   d e^6 &=& e^{34} , 
\eea 
or equivalently, defining the one--forms $z^1 = e^1 + i e^2$, $z^2 =
e^3 + i e^4$, $z^3 = e^5 + i e^6$, 
\be 
  d z^3 = \frac{1}{2} \left( i z^{1\overline{1}} - z^{2\overline{2}}
  \right) . 
\ee 
A direct check shows that 
\bea 
   dJ &=& e^{126} - e^{345} , \\
   \W_3 &=&
   \frac{1+i}{8}\left(z^{1\overline{1}}-z^{2\overline{2}}\right)\,z^3
   , \\
   \W_4 &=& \frac{1}{2}\left( e^6 - e^5 \right) . 
\eea
The Bianchi identities for the two forms imply they are closed. We
choose 
\bea 
   \tilde{K}^{(1,1)_0} &=& \beta_1 z^{1\overline{2}} +
   \overline{\beta}_1 z^{\overline{1}2} -iB_2 \left(
   z^{1\overline{1}}-z^{2\overline{2}} \right) , \\
   L &=& \frac{1+i}{4} z^{\overline{1}\overline{2}} + \alpha z^{12} , 
\eea 
where the constants $\beta_1$, $\alpha$ are complex and $B_2$ real. We also
 choose the form of $\o$ to be 
\be 
  \o = C_1 e^1 + C_2 e^2 + C_3 e^3 + C_4 e^4 + C_5 e^5 + C_6 e^6,   
\ee
with the $C_i$ real. 
Then the Bianchi identity for $G$ gives $\beta_1 =0$, $B_2 = 1/4 \, (C_6
-C_5)$. Lastly, the dilaton and four--form field equations are
equivalent to 
\bea 
   \frac{1}{8} + |\alpha|^2 &=& \frac{1}{8} (C_5 - C_6)^2 , \\
   |\alpha|^2 &=& \frac{3}{8} - \frac{1}{8} (C_5 - C_6)^2 , 
\eea 
with solution $|C_6 - C_5| = \sqrt{2}$, $|\alpha| =
1/2\sqrt{2}$. Taking for example $C_6 > C_5$ and $\alpha =
\frac{1+i}{4} e^{i\theta}$  we get the final form 
\bea 
   K &=& - \left( C_5 + \frac{1}{\sqrt{2}} \right) \left( e^{12} +
   e^{34} \right) , \\ 
   L &=& \frac{1+i}{4} \left[ (1+e^{i\theta})(e^{13}-e^{24})
     -i(1-e^{i\theta})(e^{23}+e^{14})\right] , \\ 
   G &=& \frac{1}{2}(dt + \o )\wedge (e^{126}-e^{345})
   -\frac{1}{2}\left[ C_5 e^{12} + (C_5 + \sqrt{2}) e^{34} \right]
     \wedge e^{56}  , \\ 
   \o &=& C_1 e^1 + C_2 e^2 + C_3 e^3 + C_4 e^4 + C_5 e^5 + (C_5 +
   \sqrt{2} ) e^6 . 
\eea 
This provides a six-parameter family of solutions.

\section{Supersymmetric solutions of the gauged theory} 
Our choice of gauge in the gauged theory was motivated by the desire
that the metric, structure and matter fields should all be preserved
along $V$. However we have seen this choice breaks manifest
$SU(2)$ covariance, and also that any timelike supersymmetric
solutions with non-zero Yang-Mills fields necessarily involve gauge
fields with electric components. These points have hampered our
efforts to obtain explicit Yang-Mills solutions, so instead we will
focus on the abelian case.  
To obtain $U(1)$ solutions, we set $L=0$. This implies the following
constraints on the torsion of the base:
\bea
\label{her}\W_1&=&\W_2=0,\\\W_4&=&-\frac{1}{2}d\log H.
\eea
Equation (\ref{her}) implies that the complex structure on the base is
integrable. Next, conformally rescaling the base according to
$g_6=H^{-1/2}\tilde{g}_6$, $J=H^{-1/2}\tilde{J}$,
$\O=H^{-3/4}\tilde{\O}$, we see that $\tilde{J}$, $\tilde{\O}$ define
a canonical complex structure on the base with metric $\tilde{g}$, and
\bea
\tilde{\W}_1&=&\tilde{\W}_2=\tilde{\W}_4=0,\\\label{w55}\tilde{\W}_5&=&-\frac{1}{4}d\log
H+d\phi-i_{A^3}J.
\eea
For the remainder of this section we will drop the tildes. So far,
this is the general $U(1)$ ansatz. However we will now restrict to
solutions of the specific form
\bea
\W_3&=&0,
\eea
together with $\phi=0$, $H=1$. Then the base is K\"{a}hler, and
$i_VG=0$. Furthermore, (\ref{w55}) 
then implies that $-gA^3_i$ is the potential for the Ricci form on the
base, ie
\be
\mr_{ij}=\frac{1}{2}R_{ijkl}J^{kl}=-gdA^3_{ij}
\end{equation}
Then the Bianchi identity for $K$ is equivalent to
\be
\mr=-\frac{1}{3}(8gh+g^2)J-g\tilde{K}^{(1,1)_0}.
\end{equation}
Since the scalar curvature of the base is $R=2J\lrcorner\mr$, we see
that for $hg\ge0$ we must choose a constant negative scalar curvature
base, and the choice of
base determines $\tilde{K}$. The forms may now be written
as
\bea
K&=&-d\o-\frac{\mr}{g},\\
G&=&-\star_6\Big(\frac{d\o}{2}+\frac{\mr}{g}+4hJ\Big)
\eea
Since both $\mr$ and $J$ are closed and coclosed in this context, the Bianchi
identity and field equation for G are
\bea
\label{bg1}d\star d\o&=&0,\\\label{bg2}2h(d\o^{(0,0)}+4d\o^{(1,1)_0})\wedge
J&=&\frac{1}{2g^2}\mr\wedge\mr-\frac{8h}{g}\mr\wedge
J\nonumber\\&-&(16h^2+4hg)J\wedge J \label{eq:to_cite}.
\eea
When the topological mass is zero, (\ref{bg2}) implies that we must
choose the base such that its Ricci form is decomposable. When
$h\neq0$, (\ref{bg2}) determines the (0,0) and $(1,1)_0$ parts of
$d\o$ in terms of the geometry of the base (note that the
$(2,0)+(0,2)$ part drops out), and we must then impose
(\ref{bg1}). Finally, the dilaton field equation reads
\be\label{bg3}
-4G\lrcorner G+K\lrcorner K+(8h-g)(16h-g)=0.
\end{equation} 
Within our restricted ansatz, the only place (2,0)+(0,2) forms can
arise is through $d\o$. As we have seen, these components drop out of
four form field equations. In fact they also drop out of (\ref{bg3}).
Furthermore since  $\star
d\o^{(2,0)+(0,2)}=d\o^{(2,0)+(0,2)}\wedge J$, any solution of our
system of equations may be deformed by the addition of an arbitrary
closed (2,0)+(0,2) form to $d\o$.

\subsection{Examples: h=0}
When the topological mass vanishes, we must choose the base such that
the Ricci form is decomposable and $R=-2g^2$. An example is
$\mathcal{M}_4\times H^2$, where $\mathcal{M}_4$ is any
hyperk\"{a}hler manifold. The base has metric
\be
ds^2=(e^1)^2+(e^2)^2+(e^3)^2+(e^4)^2+\frac{1}{g^2}(d\theta^2+\sinh^2\theta
d\psi^2), 
\end{equation}
and the Ricci form is $\mr=-\sinh\theta d\theta\wedge d\psi\equiv
-g^{2}e^5\wedge e^6$.
We may solve (\ref{bg1}) by taking 
\be
\o=\a\cosh\theta d\psi.
\end{equation}
Then (\ref{bg3}) implies that
\be
\a=g^{-1}.
\end{equation}
By rescaling $t$, we may express the full solution as
\bea
ds^2&=&\frac{1}{g^2}\Big(-(\Sigma^3)^2+(\Sigma^1)^2+(\Sigma^2)^2\Big)+ds^2(\mathcal{M}_4),\\G&=&\frac{g}{2}e^{1234},\;\;\;K=0,
\eea
where $\Sigma^i$ are the invariant one forms on $AdS_3$ (see appendix
A for details). 

A second example of a solution with this base may be obtained by
taking $d\o=gJ$, and the metric is
\be
ds^2=-(dt+g\rho+g^{-1}\cosh\theta
d\psi)^2+ds^2(\mathcal{M}_4)+g^{-2}(d\theta^2+\sinh^2\theta d\psi^2),
\end{equation}
where $\rho$ is the K\"{a}hler form potential on $\mathcal{M}_4$,
$d\rho=J_4$. These solutions have closed timelike curves irrespective
of the choice of $\mathcal{M}_4$, as may be
seen from the norm of $\frac{\pa}{\pa\psi}$.

\subsection{Examples: $hg>0$}
When the topological mass is nonzero, we have a wider range of allowed
bases than when $h=0$; in particular, the restrictive condition
$\mr\wedge\mr=0$ is lifted. We will now consider some examples.

\subsection*{Vacuum solutions}
Imposing $K=G=0$ on our ansatz, we may deduce that 
$d\o=-g^{-1}\mr=8hJ$, and hence $16h=g$. We must thus choose an
Einstein base with scalar curvature $R=-3g^2$, and taking
$d\o=\frac{g}{2}J$ implies that the seven dimensional solution is also
Einstein. For instance, it is well known \cite{Gary} that all 
$2n+1$-dimensional AdS spaces may be obtained as circle bundles over a
complex
hyperbolic $n$-space $\mathcal{H}^n_{\mathbb{C}}$ equipped with its Bergman metric. Explicitly, we
choose the base with metric
\be
ds^2=\frac{4}{g^2}\Big[dr^2+\frac{1}{4}\sinh^2r(\Sigma^3-\s^3)^2+\cosh^2(\frac{r}{2})\Big((\Sigma^1)^2+(\Sigma^2)^2\Big)+\sinh^2(\frac{r}{2})\Big((\s^1)^2+(\s^2)^2\Big)\Big],
\end{equation}
and we then find that
\be
(dt+\o)=\frac{1}{g}(\Sigma^3+\s^3+\cosh r(\Sigma^3-\s^3)),
\end{equation}
thus obtaining a metric on $AdS_7$. In a similar fashion we may obtain other seven dimensional Einstein
manifolds admitting Killing spinors. Some examples are circle bundles
over $H^2\times H^2\times H^2$ or $\mathcal{H}^2_{\mathbb{C}}\times
H^2$, with metrics
\bea
ds^2&=&\frac{1}{g^2}\Big[-(dt+\sum_1^3\cosh
  r_id\phi_i)^2+2\sum_1^3(dr_i^2+\sinh^2r_id\phi_i^2)\Big],\\ds^2&=&\frac{1}{g^2}\Big[-\frac{1}{4}(dt-3\sinh^2r\s^3+2\cosh\theta d\phi)^2+2\Big(d\theta^2+\sinh^2\theta d\phi\Big)\nonumber\\&+&\frac{4}{3}\Big(dr^2+\frac{1}{4}\sinh^2r((\s^1)^2+(\s^2)^2+\cosh^2r(\s^3)^2)\Big)\Big].
\eea
These both have closed timelike curves, which in contrast to $AdS$ may
not be eliminated by going to the covering space. Such solutions have
been discussed previously in eg. \cite{strom}. 

\subsection*{Squashed $AdS_7$}

Since $\mathcal{H}^3_{\mathbb{C}}$ equipped with its Bergman metric is
Einstein, we have $\tilde{K}^{(1,1)_0}=0$, and so from \eref{eq:to_cite} 
$d\o^{(1,1)_0}=0$ when we choose this base. Allowing for non-zero
fluxes on the base, the (0,0) part of $d\o$ will change to
give a squashed fibration.

To see this explicitly, consider $d\o^{(2,0)+(0,2)}=0$, $d\o\sim J$,
and for simplicity, $K=0$. We obtain a solution
provided that $4h=g$, $\mr=-g^2J=-gd\o$, and $G=-1/4J\wedge J$. In
terms of the $AdS$ siebenbeins $e^0=g^{-1}(\Sigma^3+\s^3+\cosh
r(\Sigma^3-\s^3))$, $e^1=2g^{-1}dr$, etc, the metric is
\be
ds^2=-(e^0)^2+\frac{1}{2}\d_{ij}e^ie^j
\end{equation}
We may similarly squash the other vacuum solutions given above.

\subsection*{Product base: $AdS_5$ and $AdS_3$ solutions}
When we choose the base to be a product of two or three K\"{a}hler
manifolds, we generically find upon solving for $\o$ that the seven
dimensional solution has closed timelike curves. However if we take
the base to be of the form $\mathcal{H}^2_{\mathbb{C}}\times
\mathcal{M}_2$ or $H^2\times\mathcal{M}_4$, it is possible to arrange
the curvatures in such a way that we can obtain $d\o\sim J_4$ or
$d\o\sim J_2$, and thus obtain $AdS_5\times\mathcal{M}_2$ or
$AdS_3\times\mathcal{M}_4$ solutions.

The simplest way to determine the $AdS_5$ solution is to
impose $G=0$, $K\sim vol(\mathcal{M}_2)$, and seek a base of the form
\be
ds^2=a^2\Big(dr^2+\frac{1}{4}\sinh^2r((\s^1)^2+(\s^2)^2+\cosh^2r(\s^3)^2)\Big)+ds^2(\mathcal{M}_2),
\end{equation}
together with $d\o=2a^{-1}J_4$. One may indeed find such a solution,
provided that $g=12h$, $a=3g^{-1}$, $\mathcal{M}_2=H^2$ with scalar
curvature $-2g^2/3$, and $K=g/3J_2$. This is one of the
$AdS_5$ solutions given by Maldacena and Nunez in \cite{carlos}. As in \cite{g5}, one may
deform the $AdS_5$, by the addition of $(2,0)+(0,2)$ terms to
$d\o$. Explicitly, one may add the terms
\be
\a d(\tanh^2r\s^1)+\b d(\tanh^2r\s^2),
\end{equation}
for constant $\a$, $\b$. 

To obtain the $AdS_3$ solutions, we take the base to be of the form
\be
ds^2=a^2(dr^2+\sinh^2rd\phi^2)+ds^2(\mathcal{M}_4).
\end{equation}
We seek solutions with $d\o=a^{-1}J_2$, and require that neither $K$
nor $G$ have components on the $H^2$. We find an
$AdS_3\times\mathcal{M}_4$ solution provided that $a=g^{-1}$, $g=12h$,
$\mathcal{M}_4$ is negative scalar curvature K\"{a}hler-Einstein with
$\mr_4=-g^2/3J_4$, and the forms are
\bea
K&=&\frac{g}{3}J_4,\\G&=&\frac{g}{12}J_4\wedge J_4.
\eea
This solution describes the $AdS$ fixed point of the near-horizon
limit of an M5 brane wrapped
on a K\"{a}hler four cycle in a Calabi-Yau four-fold \cite{dan}.

\section{Conclusions}
In this work we have studied bosonic field configurations of minimal
and $SU(2)$ gauged $d=7$ supergravity admitting timelike Killing
spinors, and shown that such a spinor is equivalent to an $SU(3)$
structure. We have given necessary and sufficient conditions for its
existence and hence obtained the most general superymmetric ansatz for
the theories. The bosonic fields are largely determined by the
structure, but the structure itself, before imposing the field
equations, is weakly constrained. We have exploited the general ansatz
to explicitly present numerous new solutions.

One of the more striking features we have found, in common with other
G-structure oriented studies of timelike Killing spinors, is the
apparent genericity of spacetimes with closed timelike curves among
supersymmetric solutions. In the exceptional case of the $AdS$ spaces,
the CTCs may be eliminated by going to the universal cover. In other
cases, such as the G\"{o}del solutions of the minimal theory, the CTCs
may be eliminated by taking a suitable quotient. Explicitly, taking
the flat base of section 4.1 to be toroidally compactified such that
\be
(x^{2n+1})^2+(x^{2n+2})^2<\a^{-2},
\end{equation}
eliminates the CTCs. More generally, it would be interesting to know
to what extent one might be able to eliminate CTCs in all metrics of
the form (\ref{gmetric}) by taking an appropriate quotient or
covering. If a class of such spacetimes exist for which this is not
possible, one would be forced to invoke some dynamical chronology
protection agent in string theory.

In this work we have only begun to explore the solutions contained in
the general ansatz, and have restricted attention to particularly
simple constructions. In the gauged theory we restricted attention to
a $U(1)$ truncation with K\"{a}hler base and constant $H$, $\phi$.
In particular, we have not explicitly given any
Yang-Mills solutions. Clearly, there is scope
for a more systematic study of restrictions of the general ansatz,
which we have not pursued here. 

It is known that the gauged theory admits Yang-Mills solutions
describing the near-horizon limits of branes wrapped on various
supersymmetric cycles. We expect these solutions to be contained in
the null class, where the G-structure defined by a single Killing
spinor is $(SU(2)\ltimes\mathbb{R}^4)\times\mathbb{R}$, which
naturally induces a $2+1+4$ split of the seven dimensional
spacetime. It will be interesting to analyse this case in more detail,
and in particular, we hope to systematically undertake a more refined
classification.   

\section*{Acknowledgements}
We would like to thank Gary Gibbons for useful discussions. M. C. is
supported by 
EPSRC, Cambridge European Trust and Fondazione Angelo Della Riccia. OC
is supported by a 
National University of Ireland Travelling Studentship, EPSRC, and a
Freyer scholarship.

\appendix

\section{Conventions}
We work in mostly plus signature. Indices in $7$ dimensions are given
by $\mu ,\nu, \dots$, 
in $6$ dimensions by $i,j, \dots$. The Dirac algebra is 
\be
\lbrace \G_\mu , \G_ \nu \rbrace = 2 g_{\mu\nu}\,. 
\ee
This tells us that in an orthonormal frame  $\G_0$ is antihermitian
and the $\G_i$ 
$(i=1,\dots,6)$ are hermitian.  Following the appendix to Chapter 1 in
\cite{Salam:fm} we have that the charge conjugation matrix $C$ satisfies
\be
\label{con1}
C^T = C\,, \qquad C^\dagger C = \bfI \,,\qquad \G_\mu ^T = - C \G_\mu
C^{-1}\,. 
\ee
We can therefore choose 
\be
\label{con2}
C=\bfI \,.
\ee
This implies that $\G_0$ is real and the $\G_i$ are imaginary.  We
will choose a representation (there are two inequivalent ones) such
that
\be
\G_0 \G_1 \G_2 \G_3 \G_4 \G_5 \G_6 = - \bfI
\,. \label{eq:representation} 
\ee
We also have the identity
\be
\G_{\alpha_1 \dots \alpha_n} = \frac{(-)^{[n/2]+1}}{(7-n)!}
\e_{\alpha_1 \dots \alpha_n \beta_1 \dots \beta_{7-n}} \G^{\beta_1
  \dots \beta_{7-n}}\,. 
\ee
We choose the orientation to be given by $\epsilon^{0123456} = +1$.

The Dirac conjugate $\bar \e_a$  of an anticommuting
spinor $\e^a$ is defined as
\be
\bar \e_a  =(\e^a)^\dagger \G_0\,, 
\ee
and we also define 
\be 
\bar \e^a = \e^{ab}\,\bar \e_b\,,
\label{dirac}
\ee
where $\e^{ab}$ is a constant antisymmetric matrix satisfying
$\e_{ab}\, \e^{bc}= -\delta_a^c$ that is used to raise and lower spinor
indices according to $\e^a \equiv \e^{ab} \e_b$, and $\e^{12}=1$.
On the other hand the symplectic-Majorana conjugate $\e^C$ of $\e$ is
defined to be
\be
\label{majorana}
(\e^C)^a = (\e^T)_b\, .  
\ee
Symplectic-Majorana spinors are those
for which (\ref{dirac}) is equal to (\ref{majorana}), namely
\be
(\e^T)^a = \bar \e^a\,. \label{eq:Majorana_cond}
\ee

Given four spinors $\epsilon_1$, $\dots$, $\epsilon_4$, the Fierz
identity is 
\be 
    \overline{\epsilon_1}\epsilon_2 \overline{\epsilon_3}\epsilon_4 = 
    \frac{1}{8} 
    \left[ \overline{\e_1}\e_4 \overline{\e_3}\e_2 + 
      \overline{\e_1}\Gamma_\mu \e_4 \overline{\e_3} \Gamma^\mu \e_2 -
    \frac{1}{2} \overline{\e_1}\Gamma_{\mu\nu} \e_4 \overline{\e_3}
    \Gamma^{\mu\nu} \e_2 - \frac{1}{3!}
 \overline{\e_1}\Gamma_{\mu\nu\rho} \e_4 \overline{\e_3}
    \Gamma^{\mu\nu\rho} \e_2 \right] \label{eq:Fierz_id} . 
\ee 
At various points in the text we make use of invariant one forms on
$S^3$ and $AdS_3$. In terms of the Euler angles $(\theta, \phi,\psi)$
the right invariant one forms are given by
\bea
\s^1&=&\sin\psi d\theta-\cos\psi\sin\theta d\phi,\\\s^2&=&\cos\psi
d\theta +\sin\psi\sin\theta d\phi,\\\s^3&=&d\psi+\cos\theta d\phi.
\eea
These obey $d\s^A=-\frac{1}{2}\e^{ABC}\s^B\wedge\s^C$, and one may
write the round $SU(2)$ invariant metric on a unit $S^3$ as
\be
ds^2=\frac{1}{4}\d_{AB}\s^A\s^B.
\end{equation}
By analytically continuing $\theta\rightarrow i\theta$, extending the
range of $\theta$ to $[0,\infty)$ and changing
the sign of the metric, one may obtain an $SL(2,\mathbb{R})$ invariant
metric on a unit $AdS_3$ as
\be\label{adsmet}
ds^2=\frac{1}{4}\eta_{AB}\Sigma^A\Sigma^B,
\end{equation}
where
\bea
\Sigma^1&=&\sin\psi d\theta-\cos\psi\sinh\theta
d\phi,\\\Sigma^2&=&\cos\psi+\sin\psi\sinh\theta
d\phi,\\\Sigma^3&=&d\psi+\cosh\theta d\phi.
\eea
The $\Sigma^A$ obey
\bea
d\Sigma^1=-\Sigma^2\wedge\Sigma^3,&&d\Sigma^2=-\Sigma^3\wedge\Sigma^1,\nonumber\\d\Sigma^3&=&\Sigma^1\wedge\Sigma^2.
\eea
The metric (\ref{adsmet}) clearly has closed timelike curves for all
constant 
$(\theta,\psi)$. These may be eliminated by going to the
universal cover: define
\bea
\psi&=&u+v,\\\phi&=&u-v,\\\theta&=&2r.
\eea
Then (\ref{adsmet}) becomes
\be
ds^2=-\cosh^2rdu^2+dr^2+\sinh^2rdv^2,
\end{equation}
and taking the ranges $-\infty<u<\infty$, $0\le v<2\pi$, we have the
familiar global metric on the universal cover of $AdS_3$.

Finally we note the following useful identity for a two form $A$ on a six
dimensional manifold equipped with an $SU(3)$ structure:
\be
A\lrcorner (J\wedge J)=4A^{(0,0)}+2A^{(2,0)+(0,2)}-2A^{(1,1)_0}
\end{equation}
By taking the dual of this equation we may deduce that
\bea
\star A^{(0,0)}&=&\frac{1}{2}A^{(0,0)}\wedge J,\\\star
A^{(2,0)+(0,2)}&=&A^{(2,0)+(0,2)}\wedge J,\\\star
A^{(1,1)_0}&=&-A^{(1,1)_0}\wedge J.
\eea

\section{Bilinears and the G-Structure} 
The standard strategy in applying G-structures to the solution of
supergravities is to assume the existence of a (globally defined)
Killing spinor. The existence of a globally defined spinor is
equivalent to the existence of a set globally defined forms,
constucted as bilinears in the spinor, which are invariant
under the isotropy group, $G$, of the spinor. This in turn implies a global
reduction of the principal frame bundle with structure group
$Spin(1,6)$, in the present context, to a subbundle with structure
group $G$. There are two maximal subgroups of $Spin(1,6)$ which leave
a spinor invariant, depending on whether the associated vector is
timelike or null. As we shall see below, in the timelike case of
interest to us here, a Killing spinor defines an $SU(3)$ structure. In
the null case, which we leave for future study, one has an
$(SU(2)\ltimes\mathbb{R}^4)\times \mathbb{R}$ structure. Here we shall
see how the $SU(3)$ structure arises. 

Thus, assume there exists a globally defined Killing spinor $\e^a$
satisfying the spinor equation
$\d\lambda^a =0$ and the Killing equation $\d\psi_\mu^a = 0$. 
We can define the following spinor bilinears
\bea
f^{(ab)} & = & \bar \e^a \e^b \label{eq:scalars} , \\ 
\e^{ab} V_\mu & = &  \bar \e^a \G_\mu \e^b ,  \\
\e^{ab} I_{\mu\nu} & = &  \bar \e^a \G_{\mu\nu} \e^b , \\
\Omega^{(ab)}_{\mu\nu\rho} & = &  \bar \e^a \G_{\mu\nu\rho} \e^b 
\,. \label{eq:3_form} 
\eea
From the reality properties of the gamma matrices and the symplectic
Majorana condition, the vector $V_\mu$ and the two-form  $I_{\mu\nu}$ are seen to be real, while instead the scalars
and the $3$--form can be rewritten as 
\begin{eqnarray} 
     f^a_{\;\; b} &=& - i g^A \, (T^A)^a_{\;\; b} , \\ 
     \Omega^a_{\;\; b} &=& - i X^A \, (T^A)^a_{\;\; b} , 
\end{eqnarray} 
with $g^A$, $X^A_{\mu\nu\rho}$, $A=1,2,3$, real. $(T^A)^a_{\;\; b}=
1/2 (\sigma^A)^a_{\;\; b}$ are generators of the $SU(2)$ Lie algebra,
$\sigma^A$ being the Pauli matrices, and obey 
\be 
    (T^A)^a_{\;\; b}\, (T^B)^b_{\;\; c} = \frac{1}{4}
\delta^{AB}\,\delta^a_c + \frac{i}{2} \epsilon^{ABC}\,(T^C)^a_{\;\; c}
. 
\ee  

One important consequence of the Fierz identity \eref{eq:Fierz_id}
is that $V_\mu$ is either timelike or null 
\be 
     V^2 = -\frac{1}{4} g^A\, g^A . 
\ee 
Here we focus on the timelike case
only.

Let us introduce coordinates adapted to our timelike Killing
  vector. We take $V=\frac{\pa}{\pa t}$, and write the general metric
  admitting a timelike Killing 
  vector as
\be\label{metric}
ds^2=-H^2(dt+\o)^2+g_{ij}dx^idx^j
\end{equation}
where $H$, $\o$ and $g_{ij}$ are independent of $t$. As a form,
$V=-H^2(dt+\o)=-He_0$.  
The chirality matrix on the base is
given by 
\be 
     \G_* = \G_1 \dots \G_6 , 
\ee 
and it is equal to $H^{-1}V^{\m}\G_{\m}=\G_0$, according to \eref{eq:representation}. The
Fierz identities for a symplectic-Majorana
spinor $\e^a$ imply the following projection: 
\be 
     \G_* \e^a = \frac{1}{H}  f^a_{\;\; b}\,\e^b .   
     \label{eq:Fierz_chiral}
\ee 
When $V$ is timelike the spacetime decomposes along a $6$--dimensional
Riemannian base and the symplectic-Majorana Killing spinor defines an $SU(3)$
structure. We can decompose the bosonic quantities according to the structure. 
In order to do this, first of all notice that in $6$
Riemannian dimensions a Weyl spinor $\eta$ of unit norm, satisfying 
\bea 
   \overline{\eta} \eta &=& 1, \label{eq:eta1}  \\ 
   \G_* \eta &=& i \eta , \label{eq:eta2} 
\eea
defines a canonical $SU(3)$ structure with a $2$--form $J$ and a
$3$--form $\Omega$ given by 
\bea 
   J_{ij} &=& i \overline{\eta} \, \G_{ij} \, \eta , \\ 
   \Omega_{ijk} &=&  \overline{\eta} \, \G_{ijk} \,
   \eta^{\star} . 
\eea
It is useful to note the projections 
\begin{eqnarray}
  \Gamma^i \eta &=& \frac{1}{2}\left( \delta^i_j
  + i J^i_{\;\;j} \right) 
  \G^j \eta   , \label{eq:proj1}  \\
  \G^{ij}\eta &=& 
  - \frac{1}{2} \overline{\Omega}^{ijk} \,\G_k \eta^*  - i
  J^{ij}\eta , 
\end{eqnarray}
that imply among other relations  
\begin{eqnarray}
J_{ij}\G^{ij}\eta&=&-6i\eta,\\\O_{ijk}\G^{ijk}\eta&=&-48\eta^{\star}
\\ 
J_{il}J_{jm} \G^{lm}\eta &=& - \G_{ij}\eta -2i J_{ij}\eta .
\eea
We may choose our basis so that $J$ and $\O$ take the standard
form
\bea
J&=&e^{12}+e^{34}+e^{45},\\\O&=&(e^1+ie^2)\wedge(e^3+ie^4)\wedge(e^5+ie^6).
\eea
Now consider a timelike symplectic-Majorana spinor $\e^a$ in
$7$--dimensions. It will admit, for each value of $a=1,2$, a chiral
decomposition over the base. We may write the most general expression
for $\e^1$ in terms of two orthogonal unit norm
Dirac spinors $\eta$, $\hat{\eta}$ of positive chirality in six
dimensions as
\be 
   \e^1 = \e^1_+ + \e^1_- = f \eta + g \left( \alpha \eta +
   \beta \hat{\eta} \right)^* , \label{eq:spinor_decomposition}
\ee 
where $f$, $g$, $\alpha$, $\beta$ are functions, $|\alpha |^2 + |\beta
|^2 =1$, and $f$ and $g$ can be taken to
be real functions because we are free to redefine $\eta$, $\hat{\eta}$
by an arbitrary phase (see \cite{Martelli_Sparks}). Imposing the
symplectic Majorana condition \eref{eq:Majorana_cond}, using
$\G_0\eta=i\eta$, $\G_0\hat{\eta}=i\hat{\eta}$ then implies
that $\e^2$ is given by
\bea 
    \e^2 &=&  i g \left( \alpha \eta +\beta \hat{\eta} \right) -i 
    f \eta^* . \label{eq:spinor_decomposition2} 
\eea
Now one can calculate the scalars $f^a_{\;\; b}$   and get in particular 
\bea 
   f^1_{\;\; 1} &=& i\, (f^2 -g^2) , \\ 
   f^1_{\;\; 2} &=& 2fg \,\alpha^* . 
\eea
Using the projection \eref{eq:Fierz_chiral} one finds the
following set of equations: 
\bea\label{popo}
  (H+f^2-g^2)g\b &=& 0, \\
  (H-g^2-f^2)g\a &=& 0,\\ 
  (H+g(1-2|\a|^2)-f^2)f &=& 0,\label{popop}\\
  fg\a\b &=& 0.
\eea
When $\b=0$, and
thus $|\a|^2=1$, there is the condition $f^2+g^2
=H$. Therefore we may set $f=H^{1/2}\cos\theta$, $g=H^{1/2}\sin\theta$. The angle
$\theta$ ranges in $[0,\pi ]$, since for $\beta=0$ in
\eref{eq:spinor_decomposition2}  a change $\theta\rightarrow \theta+\pi$ can be
reabsorbed by redefining $\eta\rightarrow -\eta$. Set also $\alpha=
e^{-i\gamma}$, then \eref{eq:spinor_decomposition2} can be rewritten
as 
\bea 
    \e^1 &=& H^{1/2}\left( \cos\theta \eta + \sin\theta e^{i\gamma}
    \eta^* 
    \right) , \label{eq:spinor_decomposition3_pre}  \\ 
    \e^2 &=&  H^{1/2} \left( i \sin\theta e^{-i\gamma} \eta - i
    \cos\theta \eta^*
    \right)    . \label{eq:spinor_decomposition3} 
\eea
Thus $\epsilon^a=H^{1/2}S^a_{\;\;b}\eta^b$, where $S\in SU(2)$ and $\eta^1=\eta$,
$\eta^2=-i\eta^{\ast}$. The $g=0$ solution of equations
(\ref{popo})-(\ref{popop}) is clearly the special case
$\theta=0$. When $\a=0$, either $g$ or $f$ are zero for non-zero $H$,
and this is again a special case of $\b=0$. Finally, there is a solution of (\ref{popo})-(\ref{popop})
with $f=0$, $g=H^{1/2}$, $|\a|^2+|\b|^2=1$. Together, $\eta$ and
$\hat{\eta}$ define an $SU(2)$ structure. However,
$\tilde{\eta}^{\ast}=(\a\eta+\b\hat{\eta})^{\ast}$ defines an
$SU(3)$ structure for which we may write the corresponding $\e^{1,2}$
in the form (\ref{eq:spinor_decomposition3}). Therefore the most
general timelike Killing spinor may be written in this form, and thus
defines an $SU(3)$ structure. Such a spinor will generically preserve
a single supersymmetry. Now noting that $i_VI=0$, $i_VX^A=g^AI$,
employing the projections satisfied by $\eta$,  we may
deduce the following form  for the bilinears: 
\bea
   g^1-ig^2 &=& 2Hi\sin2\theta \label{eq:decomp_scalars}
   e^{i\gamma},\:\;g^3=-2H\cos2\theta,\\
   I &=& HJ,\\ 
   \label{form}X^1-iX^2 &=& -H^{-1}(g^1-ig^2)V\wedge
J+2Hi(\sin^2\theta e^{2i\gamma}\O-\cos^2\theta\overline{\O}), \\ 
   X^3 &=& -H^{-1}g^3V\wedge
   J-H\sin2\theta(e^{i\gamma}\O+c.c.). \label{eq:decomp_X3} 
\eea

\section{Necessary and sufficient conditions for supersymmetry
   \label{app:nec_and_suff}} 
The type of structure defined by the Killing spinor is determined by
its intrinsic torsion. We have an $SU(3)$ structure in seven
dimensions, specified by $(V,J,\O)$, with an associated six dimensional
structure, specified by $(J,\O)$, or equivalently by the chiral unit norm
spinor $\eta$. Given such a six dimensional structure, there is no
obstruction to finding a connection $\nabla^\prime$ that preserves it,
$\n^\prime\eta 
=0$. $\n^\prime$
is not unique, and different inequivalent classes of structure
preserving connections are parametrized by the
part of the torsion tensor called the intrinsic torsion. In
\cite{ChiossiSalamon} it is shown that for an $SU(3)$ structure in $6$
dimensions there are $5$ modules of the intrinsic torsion given by 
\bea 
  d\O^{(2,2)} &=& \W_1 \, J\wedge J + \W_2 \, J , \\  
  \W_3 &=& (d\, J)^{(2,1)_0} , \\ 
  \W_4 &=& \frac{1}{2} \, J \lrcorner d J , \\ 
  \W_5 &=& \frac{1}{2} Re\, \O \lrcorner \, d( Re\, \O ) . 
\eea
The intrinsic torsion may be calculated in terms of the metric and
matter fields by
applying the Killing spinor equation to the spinor
bilinears. Furthermore, the
vanising of $\d\lambda$ relates various components of the bosonic
fields to each other. In this appendix we will determine
the constraints on the bosonic fields of the minimal and gauged
theories implied by the existence a timelike Killing spinor. As we
shall see, most of the field content of the theories is determined by
the structure. As a final step, we will show that the constraints we
derive on the bosonic fields are also sufficient to ensure the
existence of a timelike Killing spinor, and thus that we have derived
the most general bosonic field configuration compatible with timelike
supersymmetry. We will first work out the constrints for the minimal
theory, and the results are then straightforwardly modified to account
for the gauging.

\subsection{Differential constraints} 
The various bispinors satisfy differential equations as a consequence of
the Killing spinor equation. Using
\be
\n_{\m}(\e^{aT}A\e^b)=(\n_{\m}\e^a)^TA\e^b+\e^{aT}A(\n_{\m}\e^b)
\end{equation}
where $A$ is any matrix in the Clifford algebra, and employing the
Killing spinor equation in the minimal theory, we may deduce
\bea
\label{dg}dg^A&=& -\frac{1}{5}\e^{ABC}F^B\lrcorner
X^C-\frac{8}{5}i_VF^A-\frac{2}{5}X^A\lrcorner
G,\\\label{dV}\n_{\m}V_{\v}&=&\frac{1}{5}\Big(3G\lrcorner\star
V+2HJ\lrcorner G+\frac{1}{2}F^A\lrcorner \star
X^A-2g^AF^A\Big)_{\m\v},\\\label{dJ}d(HJ)_{\m\v\s}&=&\frac{1}{5}\Big(
3F^A_{\a[\m}X^{A\;\;\;\;\;\a}_{\;\;\;\v\s]}-2HG_{\a\b\gamma[\m}\star
  J_{\v\s]}^{\;\;\;\;\a\b\gamma}+6i_VG_{\m\v\s}\Big),\\\label{dX}
  dX^A_{\m\v\s\t}&=&\frac{1}{5}\Big(8F^A\lrcorner\star
  V_{\m\v\s\t}+12HF^A\wedge J_{\m\v\s\t}+4\e^{ABC}F^B_{\a[\m}\star
      X^{C\;\;\;\;\a}_{\v\s\t]}\nonumber\\&-&8g^AG_{\m\v\s\t}+6G_{[\m\v}^{\;\;\;\;\;\;\a\b}\star X^A_{\s\t]\a\b}\Big).
\eea
where $i_VA$ denotes the vector $V$ contracted on the first index of
the form $A$. Note that (\ref{dV}) implies that $V$ is Killing.
Next, by successively contracting $\d\lambda=0$ with $\e^{aT}$,
$\e^{aT}\G_{\m}$,..., $\e^{aT}\G_{\m\v\s}$, and splitting the
symplectic Majorana indices into symmetric and and antisymmetric
parts, we find (among others) the following constraints:
\bea
\label{dl1}g^A\pa_{\m}\phi&=&=\frac{1}{5}\Big(-2i_VF^A_{\m}+2X^A\lrcorner
G_{\m}+\e^{ABC}F^B_{\a\b}\lrcorner X^{C\a\b}_{\m}\Big) , \\\label{dl2}(d\phi\wedge
V)_{\m\v}&=&\frac{1}{5}\Big(2G\lrcorner\star
V-2HJ\lrcorner G-\frac{1}{2}F^A\lrcorner \star
X^A-\frac{1}{2}g^AF^A\Big)_{\m\v},\\
\label{dl3}2H(d\phi\wedge J)_{\m\v\s}&=&\frac{1}{5}
\Big(3F^A_{\a[\m}X^{A\;\;\;\;\;\a}_{\;\;\;\v\s]}-2HG_{\a\b\gamma[\m}\star
  J_{\v\s]}^{\;\;\;\;\a\b\gamma}-4i_VG_{\m\v\s}\Big),\\\label{dl4}(d\phi\wedge X^A)_{\m\v\s\t}&=&\frac{1}{5}\Big(-2F^A\lrcorner\star
  V_{\m\v\s\t}+2HF^A\wedge J_{\m\v\s\t}+4\e^{ABC}F^B_{\a[\m}\star
      X^{C\;\;\;\;\a}_{\v\s\t]}\nonumber\\&+&2g^AG_{\m\v\s\t}
+6G_{[\m\v}^{\;\;\;\;\;\;\a\b}\star X^A_{\s\t]\a\b}\Big).
\eea
Combining (\ref{dg}) with (\ref{dl1}) we obtain
\be
i_VF^A=-\frac{e^{-\phi}}{2}d(g^Ae^{\phi}).
\end{equation}
Given the Bianchi identity for $F$, and, as we will show below, that
$\mathcal{L}_V\phi=0$, this implies that
\be
\mathcal{L}_VF^a=\mathcal{L}_Vg^A=0.
\end{equation}
Next (\ref{dV}) and (\ref{dl2}) combine to give
\be\label{jjj}
e^{-2\phi}d(e^{2\phi}V)=2i_V\star G-g^AF^A.
\end{equation}  
We also find
\be
\frac{e^{2\phi}}{2}d(He^{-2\phi}J)=i_VG \label{eq:dJ}
\end{equation}
which (given the Bianchi identity for $G$, and $\mathcal{L}_VH=0$)
implies that
\be
\mathcal{L}_VG=\mathcal{L}_VJ=0.
\end{equation}
Finally, (\ref{dX}) and (\ref{dl4}) give
\be
\frac{e^{\phi}}{2}d(e^{-\phi}X^A)=F^A\lrcorner\star V+HF^A\wedge J-g^AG.
\end{equation}
Given the form (\ref{form}), \eref{eq:decomp_X3} of the $X^A$, we may
deduce that  
\be
\mathcal{L}_V\O=0.
\end{equation}
Therefore $V$ generates a symmetry not just of the metric and matter
fields but also of the G-structure.

\subsection{Constraints from $\delta\lambda^a=0$}
In order to deduce the constraints on the bosonic fields of the theory
implied by the vanishing of $\d\lambda^a$, rather than solving
(\ref{dl1})-(\ref{dl4}) directly, it is technically much more
convenient to break the manifest global $SU(2)$ symmetry of the
theory, and to work with the Dirac spinor $\eta$. To this end, let us define
\bea
\tilde{F}^a_{\;\;\;b}&=&(S^{-1}FS)^a_{\;\;\;b},\\K&=&\tilde{F}^3,\\L&=&\tilde{F}^1-i\tilde{F}^2.
\eea
Then $\d\lambda^a=0$ implies the following pair of equations:
\bea
(\frac{5}{2}\pa_{\m}\phi\G^{\m}+\frac{i}{4}K_{\m\v}\G^{\m\v}-\frac{1}{6}\star
G_{\m\v\s}\G^{\m\v\s})\eta+\frac{1}{4}L_{\m\v}\G^{\m\v}\eta^{\star}&=&0,\\(\frac{5}{2}\pa_{\m}\phi\G^{\m}-\frac{i}{4}K_{\m\v}\G^{\m\v}-\frac{1}{6}\star
 G_{\m\v\s}\G^{\m\v\s})\eta^{\star}+\frac{1}{4}\overline{L}_{\m\v}\G^{\m\v}\eta&=&0. 
\eea
Taking the complex conjugate of the second equation, and employing the
reality properties of the gamma matrices together with the projection
$\G_0\eta=i\eta$, we may rewrite these equations in the following form:
\bea
(-5\pa_0\phi +\frac{1}{2}K_{ij}\G^{ij}+\star
G_{0ij}\G^{ij})\eta-L_{0i}\G^i\eta^{\star}&=&0,\\
(5\pa_i\phi\G^i-K_{0i}\G^i-\frac{1}{3}\star
G_{ijk}\G^{ijk})\eta+\frac{1}{2}L_{ij}\G^{ij}\eta^{\star}&=&0.
\eea
What we have done is to split the supersymmetry variation of $\d\lambda$
into positive and negative chirality parts on the base. To solve these
equations we decompose the above forms into $SU(3)$
irreducible representations. Successively contracting with $\eta^T$,
$\overline{\eta}$, 
$\eta^T\G_i$,..., $\overline{\eta}\G^{ijk}$ we deduce that
$\delta\lambda^a=0$ is equivalent to
\bea
\label{phi}\pa_0\phi&=&0,\\(\frac{1}{2}K_{ij}+\star
G_{0ij})^{(2,0)+(0,2)+(0,0)}&=&-\frac{1}{8}(L_{0k}\O^k_{\;\;ij}+c.c.),\\\frac{1}{3}\O_{ijk}\star
G^{ijk}&=&\frac{i}{2}\overline{L}_{ij}J^{ij},\\
\label{p} \frac{1}{2}(\d_i^j-iJ_i^{\;\;j})  (5\pa_j\phi-K_{0j}+i\star  
G_{jkl}J^{kl})&=&-\frac{1}{4}L_{jk}\O_i^{\;\;jk}. 
\eea
The primitive forms $\star G_{ijk}^{(2,1)_0+(1,2)_0}$, $\star
G_{0ij}^{(1,1)_0}$, $K_{ij}^{(1,1)_0}$ and $L_{ij}^{(1,1)_0}$,
traceless with respect to $J$, drop out
  of the supersymmetry variation and are unconstrained here.

Now we will re-express the differential constraints on the
structure derived in the last subsection in terms of the transformed
two forms. Let us define the rotated quantities
\bea
\tilde{g}^a_{\;\;\;b}&=&(S^{-1}gS)_{\;\;\;b}^a,\\\tilde{X}_{\;\;\;b}^a
&=&(S^{-1}XS)_{\;\;\;b}^a.
\eea
They take the form
\bea
\tilde{g}^1-i\tilde{g}^2&=&0,\\\tilde{g}^3&=&-2H,\\\tilde{X}^1-i\tilde{X}^2
&=&-2iH\overline{\O},\\\tilde{X}^3&=&2V\wedge
J.
\eea
Also define
\be
C_{\m }\,^a_{\;\;\;b}=i(S^{-1})^a_{\;\;\;c}\,\pa_{\m}S^{c}_{\;\;\;b}. 
\label{eq:C_def}
\end{equation}
 Then we find
\bea
\label{iii}K_{0\m}&=&\frac{e^{-\phi}}{H}\pa_{\m}(He^{\phi}),\\L_{0\m}&=&i(C^1-iC^2),\\
\frac{e^{-2\phi}}{2H}d(e^{2\phi}V)_{ij}&=&
\star
G_{0ij}+K_{ij},\\\label{pp}\frac{e^{2\phi}}{2H}d(He^{-2\phi}J)_{ijk}&=&
 G_{0ijk}, 
\eea 
\be 
\label{eq:dO}i\star_6( \frac{e^\Phi}{H}
d(e^{-\Phi}H\O)+iC^3\wedge\O)_{ij}=\overline{L}_{ij}^{(0,0)}-
2\overline{L}_{ij}^{(1,1)_0},
\ee
where $\star_6$ denotes the Hodge dual on the base. Now, we can calculate all modules of the intrinsic torsion
of the $SU(3)$ structure on the base using (\ref{pp}), \eref{eq:dO},
obtaining the equations quoted in \eref{eq:W1}-\eref{eq:W5}. These,
together with equations (\ref{phi})-(\ref{p}) and
(\ref{iii})-\eref{eq:dO} (which are rewritten as in
\eref{eq:ns_phi}-\eref{eq:ns_G})
are necessary conditions for supersymmetry. In fact they are also
sufficient, as we now show.

\subsection{Sufficient conditions for supersymmetry in the minimal theory}
The necessary conditions we have obtained gaurantee the vanishing of
$\d\lambda$, as may be seen by substituting (\ref{phi})-(\ref{p}) back
in and employing the projections satisfied by $\eta$. It remains to be
verified that a solution of the Killing spinor equation always
exists. Rewriting the Killing spinor equation in terms
of $\eta$ we get 
\bea 
  D_\mu \eta^a &+& \frac{1}{2}\pa_\mu (\log H)\eta^a + C^A_\mu (
  T^A)^a_{\;\;b} \eta^b -\frac{i}{10}\, F_{\l_1\l_2\;\;b}^{\;\;\;\;\;\;\;\;a}
  \,  \left(\G_\mu^{\;\;\l_1\l_2} -
  8\delta^{l_1}_\mu \G^{\l_2} \right) \eta^a \nonumber \\ &+& \frac{1}{80}
  G_{\l_1\dots\l_4}\left(\G_\mu^{\;\;\l_1\dots\l_4}
  -\frac{8}{3}\delta^{\l_1}_\mu \G^{\l_2\l_3\l_4} \right) \eta^a = 0. 
\eea
These two equations are not independent as can be seen by 
taking the complex conjugation of the second one. 
Separating the negative and the positive chirality parts we get two
different equations: one is algebraic and the other differential. The
first one reads 
\bea 
 0 &= & i  (C^1 + i C^2) \eta^* -\frac{1}{4}\omega_{\m
  \lambda_1 \lambda_2}(\G
-\G^*)^{\lambda_1 \lambda_2}\eta  \nonumber \\  
 &+& \frac{i}{20}\, \left\{ K_{\l_1\l_2} \left[ \left(\G -\G^*
   \right)_\mu^{\;\;\l_1\l_2} -
  8\delta^{l_1}_\mu \left( \G -\G^* \right)^{\l_2} \right] \eta
    - i  \overline{L}_{\l_1\l_2} \left[ \left(\G +\G^*
   \right)_\mu^{\;\;\l_1\l_2} \right. \right. \nonumber \\   &-&
   \left. \left. \hspace{-.2cm}
  8\delta^{l_1}_\mu \left( \G +\G^* \right)^{\l_2}
  \right]\hspace{-.1cm} \eta^*
  \right\} \hspace{-.1cm} -\hspace{-.1cm} \frac{1}{80} \hspace{-.1cm}
  G_{\l_1\dots\l_4}\hspace{-.1cm}\left[  \left(\G\hspace{-.1cm}
    -\G^*\right)_\mu^{\;\;\l_1\dots\l_4}
  \hspace{-.1cm}-\frac{8}{3}\delta^{\l_1}_\mu \left(\G -\G^*\right)^{\l_2\l_3\l_4}
  \hspace{-.1cm}\right] \hspace{-.1cm} \eta   . \label{eq:algebraic} 
\eea 
Nothing that (\ref{jjj}) implies the
following expressions for the spin connection,
\bea
\o_{0i0}&=&K_{0i}-\pa_i\phi,\\\o_{ij0}&=&\o_{0ij}=\star
G_{0ij}+K_{ij},
\eea
one may verify that the algebraic equation \eref{eq:algebraic} is
satisfied. Next, the remaining differential equation for $\eta$ is
given by  
\bea 
  & & (\partial_\mu + \frac{1}{4}\omega_{\mu ij}\G^{ij}) \eta  + \pa_\mu (\log
  H)\eta + \frac{1}{2} C^3_\mu \eta  
- \frac{i}{40}\, \left\{ K_{\l_1\l_2} \left[ \left(\G +\G^*
   \right)_\mu^{\;\;\l_1\l_2} \right. \right. \nonumber \\ & - &
   \left. \left.
  8\delta^{l_1}_\mu \left( \G +\G^* \right)^{\l_2} \right] \eta -i 
 \overline{L}_{\l_1\l_2} \left[ \left(\G -\G^*
   \right)_\mu^{\;\;\l_1\l_2} -
  8\delta^{l_1}_\mu \left( \G -\G^* \right)^{\l_2} \right] \eta^*
 \right\} \nonumber \\ &+& \frac{1}{160}
  G_{\l_1\dots\l_4}\left[  \left(\G +\G^*\right)_\mu^{\;\;\l_1\dots\l_4}
  -\frac{8}{3}\delta^{\l_1}_\mu \left(\G +\G^*\right)^{\l_2\l_3\l_4}
  \right] \eta = 0 .  
\eea 
The $\mu = 0$ component reduces to 
\be
\pa_0\eta=0 , 
\end{equation}
and is satisfied by any time-independent spinor satisfying the
required projections. Next, consider the $\mu =i$ components. A
straightforward but long calculation shows that it can be rewritten as
\bea 
 \left[ \right. &&  \hspace{-.5cm} \n_m  + \frac{1}{4}\W_4^a \,
   \G_{am}+ 
   \frac{i}{4} \left( 3 \W_4 + 2 \W_5 \right)\,^r \, J_{rm} - 
    \frac{1}{32}   \W_1^* \,
    \O_{ma_1a_2} \G^{a_1a_2} -
    \frac{i}{8} \W_{3ma_1a_2} 
  \G^{a_1a_2} \nonumber \\ &+& \left.  
   \frac{i}{32}  \W^*_{2mj}\,\O^j_{\;\;a_1a_2}  \G^{a_1a_2} \right]
 \eta = 0 
\eea
As we show in  appendix \ref{app:contorsion}, this is the most general $SU(3)$
preserving connection on the base. Therefore we always have a solution
of the Killing spinor equation, and we have derived necessary and
sufficient conditions for supersymmetry.

\subsection{Necessary and sufficient conditions for supersymmetry in the
  gauged theory}
Having determined the constraints for supersymmetry in the minimal
theory, it is a straightforward matter to modify the results to
account for the gauging. We will therefore only briefly quote our
results. However there is one point which deserves mention. A
desirable feature we want to maintain in the gauged theory is that the
Killing vector $V$ generates a symmetry not just of the metric and
matter fields but also of the G-structure. The differential
constraints on the structure in the gauged theory are
\bea
i_VF^A&=&-\frac{1}{2}e^{-\phi}(d(e^{\phi}g^A)+g\e^{ABC}A^Be^{\phi}g^C),
\\e^{-2\phi}d(e^{2\phi}V)&=&2i_V\star G-g^AF^A-8Hhe^{-4\phi}J,
\\e^{2\phi}d(He^{-2\phi}J)&=&2i_VG,
\\e^{\phi}d(e^{-\phi}X^A)\hspace{-.075cm} +\hspace{-.075cm}
g\e^{ABC}A^B\hspace{-.075cm} \wedge
X^C\hspace{-.1cm}&=&\hspace{-.075cm} 2F^A\lrcorner\star
V \hspace{-.075cm} + \hspace{-.075cm} 2HF^A \hspace{-.075cm} \wedge J
-\hspace{-.075cm} 2g^AG+g\star X^A, 
\eea
and $V$ is again Killing. As before, from $\d\lambda=0$, we have
$i_Vd\phi=0$. Therefore, imposing the gauge
\be
i_VA^A=\frac{1}{2}e^{\phi}g^A, \label{eq:gauge1}
\end{equation}
ensures that (given the Bianchi identities) the matter fields and the
structure are also preserved along $V$. However, (\ref{eq:gauge1}) does not
entirely fix the gauge freedom; we may still perform time independent
gauge transformations, under which both $A^A_0$ and $g^A$ transform in
the adjoint. We may thus eliminate the scalars $\theta$, $\gamma$
altogether by imposing the gauge
\be\label{eq:gauge2}
A^1_0=A^2_0=0,\;\;\;A^3_0=-e^{\phi},
\end{equation}
and the structure simplifies to
\bea
g^1&=&g^2=0,\\g^3&=&-2H,\\X^1-iX^2&=&-2iH\overline{\O},\\X^3&=&2V\wedge
J.
\eea
It is now a simple matter to modify the results of the minimal
theory. What we find in the gauge (\ref{eq:gauge1}), \eref{eq:gauge2} is reported in
eqs.\eref{eq:ns_phi_g}-\eref{eq:W5_g}.

\section{Intrinsic torsion \label{app:contorsion}} 
In this appendix we derive a formula for the covariant derivative that
leaves invariant the $SU(3)$ structure. The procedure to do it is well
understood and we follow the lines of \cite{Gauntlett_Pakis}.

Write the connection as $\G^i_{\;\; jk} = C^i_{\;\; jk} + K^i_{\;\;
  jk}$, where $C^i_{\;\; jk}$ are the Christoffel symbols and $K_{ijk}
= K_{[i|j|k]}$ is the contorsion tensor. The contorsion is equivalent
to the torsion $T^i_{\;\; jk}$  in that it satisfies 
\bea 
  T^i_{\;\; jk} &=& 2 K^i_{\;\; [jk]} , \\ 
  K^i_{\;\; jk} &=& \frac{1}{2} \left( T^i_{\;\; jk} +
  T^{\;\;i}_{j\;\;\;\;k} + T^{\;\;i}_{k\;\;\;\;j} \right) . 
\eea
The contorsion is a tensor in $T^* \otimes so(6) \simeq ( T^* \otimes
su(3) ) \oplus ( T^* \otimes su(3)^\perp )$. $T^* \otimes
su(3) )$ decomposes under $SU(3)$ as 
\be 
( \mathbf{3} + \overline{\mathbf{3}} ) \times \mathbf{8} =
(\mathbf{15} + \overline{\mathbf{15}} ) + (\mathbf{6} +
\overline{\mathbf{6}} ) + ( \mathbf{3} + \overline{\mathbf{3}} ) , 
\ee 
while $T^* \otimes su(3)^\perp$ as 
\be 
( \mathbf{3} + \overline{\mathbf{3}} ) \times ( \mathbf{3} +
\overline{\mathbf{3}} +  \mathbf{1} ) = (  \mathbf{8} +
\mathbf{8}^\prime ) + ( \mathbf{6} + \overline{\mathbf{6}} ) + (
\mathbf{3} + \overline{\mathbf{3}} ) + ( \mathbf{3}^\prime +
\overline{\mathbf{3}^\prime} ) + ( \mathbf{1} + \mathbf{1}^\prime )  . 
\ee 
When acting on $SU(5)$ invariants, only this latter part of the
contorsion contributes. Now we rewrite a general contorsion tensor
according to its $SU(3)$ decomposition as 
\bea 
  K_{lmn} &=&\left( L^{(1)}_m \, J_{ln} + L^{(2)}_{[l} \, J_{n]m}
  +L^{(3)}_{[l} \, g_{n]m} \right) + ( k \O_{lmn} + c.c. ) + (
  f_{mj}\O^j_{\;\;ln} + c.c ) \nonumber \\ &+& (T^{(1)}_{mk}\,\Ob^k_{\;\;ln} +
  T^{(2)}_{[l|j|} \,\Ob^j_{\;\;n]m} + c.c. ) + (U_{[ln]m} + c.c.) , 
\eea 
where $K\in\mathbb{C}$ and,  in complex notation, $f_{lm}=
f_{\lambda\overline{\mu}} + f_{\overline{\lambda}\m} \in \mathbb{R}$,
$f_{\lambda\overline{\mu}} J^{\lambda\overline{\mu}} =0$, $T_{lm} =
T_{(\lambda \m )}$, $U_{lmn}=U_{\overline{\l}(\m\v)}$,
$U_{\overline{\l}\m\v} J^{\overline{\l}\m} = 0$. Suppose that
$K_{lmn}$ is such $\n^\prime J = 0 = \n^\prime \O$, then the exterior
derivative of $J$, $\O$ is given by 
\bea 
  \frac{1}{6} dJ_{i_1 i_2 i_3} &=& K^r_{\;\;[i_1i_2}\,J_{|r|i_3]} , \\ 
  \frac{1}{12} d\O_{i_1 \dots i_4} &=& K^r_{\;\;[i_1i_2}\,\O_{|r|i_3
      i_4]} . 
\eea 
From these we calculate the intrinsic torsion modules and find 
\bea 
  \W_1 &=& 8 k^* , \\ 
  \W_2 &=& 8i f^* , \\ 
  \W_{3 i_1i_2i_3} &=& 6i T^{(1)*}_{[i_1|j|}\Omega^j_{\;\;i_2i_3]} , \\ 
  \W_4 &=& L^{(3)} - L^{(2)}\lrcorner \, J , \\ 
  \W_5 &=& - \frac{3}{2} L^{(3)} + 3 \left( L^{(1)} + \frac{1}{6}
  L^{(2)} \right) \lrcorner\, J . 
\eea 
Some components of the contorsion are not determined by the $\W_i$ and
are those under which the structure is preserved, corresponding to the
freedom in choosing an $SU(3)$ preserving connection. As a matter of
fact we can rewrite the contorsion now as 
\bea 
  K_{lmn} &=&\left( \frac{3}{2} \W_4 + \W_5 \right)\,^r \, J_{r[l} \,
    J_{n]m} - \left( \frac{1}{2} \W_4 + \W_5 \right)\,_{[l} \, g_{n]m} 
  + \frac{1}{8}\left( \W_1^* \O_{lmn} + c.c. \right)  \nonumber \\ &-& 
   \left(
  \frac{i}{2} W^*_{3lmn} + c.c \right) + 
  \left( \frac{i}{8}\W^*_{2mj}\,\O^j_{\;\;ln} + c.c. \right)
  \nonumber \\ 
   &+& \left( L^{(1)}_m \, J_{ln} + 3 L^{(1)}_{[l}\, J_{n]m} + 3
   L^{(1)r} \, J_{r[l}\,g_{n]m} \right) \nonumber \\ &+& \left[ \left(
     T^{(2)}-2T^{(1)}\right)\,_{[l|r|}\Ob^r_{\;\;n]m} + c.c. \right]
    + (U_{[ln]m} + c.c.) .  
\eea 
One can directly check that the last two lines leave both $J$ and $\O$
invariant or, equivalently, $\eta$. The first two lines therefore
define the intrinsic contorsion $K^0_{lmn}$. Notice also that the
combination $\frac{3}{2} \W_4 + \W_5$ is conformally invariant
\cite{ChiossiSalamon}. 
The last equation we need
is the form assumed by $\n^\prime\eta=0$ which concretely reads 
\bea 
 \left[ \right. && \hspace{-.5cm} \n_m  + \frac{1}{8} \left( 3 \W_4 + 2 \W_5 \right)\,^r \, J_{ra_1} \,
    J_{a_2 m} \G^{a_1a_2} - \frac{1}{8} \left( \W_4 + 2 \W_5
    \right)\,_{a_1} \G^{a_1}_{\;\;m}  \nonumber \\ &-& 
    \frac{1}{32}  \left( \W_1^* 
    \O_{ma_1a_2} + \W_1 \Ob_{ma_1a_2} \right) \G^{a_1a_2} 
  +  \frac{i}{8} \left( \W^*_{3ma_1a_2} - \W_{3ma_1a_2} \right)
  \G^{a_1a_2} \nonumber \\ &+& \left. 
   \frac{i}{32} \left( \W^*_{2mj}\,\O^j_{\;\;a_1a_2} -
   \W_{2mj}\,\Ob^j_{\;\;a_1a_2} \right) \G^{a_1a_2} \right] \eta = 0
  \label{eq:covariant_der} . 
\eea 
Using the projection \eref{eq:proj1} and the following we see that
\eref{eq:covariant_der} can be rewritten as 
\bea 
 \left[ \right. &&  \hspace{-.5cm} \n_m  + \frac{1}{4}\W_4^a \,
   \G_{am}+ 
   \frac{i}{4} \left( 3 \W_4 + 2 \W_5 \right)\,^r \, J_{rm} - 
    \frac{1}{32}   \W_1^* \,
    \O_{ma_1a_2} \G^{a_1a_2} -
    \frac{i}{8} \W_{3ma_1a_2} 
  \G^{a_1a_2} \nonumber \\ &+& \left.  
   \frac{i}{32}  \W^*_{2mj}\,\O^j_{\;\;a_1a_2}  \G^{a_1a_2} \right]
 \eta = 0 
\label{eq:fundamental} 
\eea

\section{Integrability Conditions \label{app:integrability}}
Let us define $\d\lambda=\Delta_{\lambda}\e$,
$\d\psi_{\m}=\mathcal{D}_{\m}\e$. Then we may obtain the following
integrability condition from commuting the Killing spinor equation
with $\d\lambda$:
\bea
\sqrt{5}\G^{\m}[\mathcal{D}_{\m},\Delta_{\lambda}]\e^a&=&\Big(
\frac{1}{2}P+
\frac{1}{6}Q_{\m\v\s}\G^{\m\v\s}+\frac{e^{2\phi}}{96}d(e^{-2\phi}G)_{\m\v\s\t\rho}\G^{\m\v\s\t\rho}\Big)\e^a\nonumber\\&+&\Big(iR^A_{\m
  }\G^{\m}+\frac{ie^{-\phi}}{6}d(e^{\phi}F^{A})_{\m\v\s}\G^{\m\v\s}\Big)T^{Aa}_{\;\;\;\;\;b}\e^b\nonumber\\&+&\sqrt{5}\Big(\frac{1}{60}G_{\m\v\s\t}\G^{\m\v\s\t}\d^a_b+\frac{3i}{5}F^{A}_{\m\v }\G^{\m\v}T^{Aa}_{\;\;\;\;\;b}\Big)\d\lambda^b,
\eea
where $P$, $Q$, $R$ are defined by
eqs.(\ref{eq:P},\ref{eq:Q},\ref{eq:R}), 
and the dilaton, four form and two form field equations are
respectively $P=0$, $Q=0$, $R^A=0$. Imposing the dilaton and four form
field equations, and the Bianchi identities for the forms, the
integrability condition reduces to
\be
R^A_{\m}\G^{\m}T^{Aa}_{\;\;\;\;\;b}\e^b=0.
\end{equation}
As in the analysis of $\d\lambda=0$, it is convenient to transform this
expression and work in terms of $\eta$. Writing $\tilde{R}=S^{-1}RS$,
and using the same procedure as before, we may deduce
\bea
\tilde{R}^3_{\m}&=&0,\\(\tilde{R}^1-i\tilde{R}^2)_0&=&0,\\(\d^j_i+iJ^{\;\;\;j}_{i})(\tilde{R}^1-i\tilde{R}^2)_j&=&0.
\eea
Next we consider the integrability condition for the Killing spinor
equation. After a long calculation we obtain
\bea
\G^{\v}[\mathcal{D}_{\m},\mathcal{D}_{\v}]\e^a&=&\Big[-\frac{1}{2}E_{\m\v}\G^{\v}+e^{2\phi}d(e^{-2\phi}G)^{\v\s\t\rho\xi}\Big(-\frac{1}{120}g_{\m\v}\G_{\s\t\rho\xi}+\frac{1}{200}\G_{\m\v\s\t\rho\xi}\Big)\nonumber\\&+&\frac{1}{10}Q^{\v\s\t}\Big(\frac{1}{2}\G_{\m\v\s\t}-g_{\m\v}\G_{\s\t}\Big)\Big]\e^a\nonumber\\&+&\Big[\frac{ie^{-\phi}}{5}d(e^{\phi}F^A)^{\v\s\t}\Big(2g_{\m\v}\G_{\s\t}+\frac{1}{6}\G_{\m\v\s\t}\Big)-\frac{i}{5}R^{A\v}(-4g_{\m\v}+\G_{\m\v})\Big]T^{Aa}_{\;\;\;\;\;b}\e^b\nonumber\\&+&\Big[\pa_{\m}\phi\d^a_b-\frac{i}{25}F^{A\v\s}(8g_{\m\v}\G_{\s}-\G_{\m\v\s})T^{Aa}_{\;\;\;\;\;b}\nonumber\\&+&\frac{1}{25}G^{\v\s\t\rho}\Big(-\frac{2}{3}g_{\m\v}\G_{\s\t\rho}+\frac{1}{4}\G_{\m\v\s\t\rho}\Big)\d^a_b\Big]\d\lambda^b=0.
\eea
Given the Bianchi identites and the field equations imposed for and
implied by the vanishing of the integrability condition for
$\d\lambda$, and converting to the dirac spinor $\eta$, this reduces
to the pair of equations
\bea
\label{E1}-\frac{1}{2}E_{\m\v}\G^{\v}\eta-\frac{1}{5}(\tilde{R}^1-i\tilde{R}^2)^j(-4g_{\m
  j}+\G_{\m j})\eta^{\star}&=&0,\\-\frac{1}{2}E_{\m\v}\G^{\v\star}\eta-\frac{1}{5}(\tilde{R}^1-i\tilde{R}^2)^j(-4g_{\m
  j}+\G_{\m j}^{\star})\eta^{\star}&=&0.
\eea
Taking the $i$ component we may deduce
\be
   E_{j0}\eta=0, 
\ee
\be
   \label{E2}-E_{ij}\G^j\eta-\frac{2}{5}(\tilde{R}^1-i\tilde{R}^2)_j 
   (-4\d^j_i+\G^{\;\;j}_{i})\eta^{\star})=0 .
\ee
Hence $E_{j0}=0$, while contracting the second expression with $\e^T$
we obtain
\be
(\tilde{R}^1-i\tilde{R}^2)_j(-\d^j_i+\frac{1}{5}(\d^i_j+iJ^{\;\;j}_i))=0.
\end{equation}
Thus $\tilde{R}^A_{\m}=0$, and so $R^A_{\m}=0$. Next contracting
(\ref{E2}) with $\overline{\eta}\G_k$ implies
\be
E_{ij}(\d^j_k\pm iJ^{\;\;j}_k)=0,
\end{equation}
so $E_{ij}=0$, and the 0 component of (\ref{E1}) then implies that
$E_{00}=0$. 

In summary, given the existence of a timelike Killing spinor, it is
sufficient to impose the Bianchi identites and the four form and
dilaton field equations. The remaining field equations are implied by
supersymmetry.  

In the gauged theory the structure of the integrability conditions is
identical. The additional terms which arise are such that one now
obtains the gauged theory field equations and Bianchi identities 
in precisely the same fashion, together with the additional terms
$\sqrt{5}(8he^{-4\phi}-ge^{\phi})\d\lambda$,
$\frac{2}{\sqrt{5}}(m+2he^{-4\phi})\d\lambda$ in
$\G^{\v}[\mathcal{D}_{\v},\Delta_{\lambda}]$,
$\G^{\v}[\mathcal{D}_{\m},\mathcal{D}_{\v}]$ respectively. Thus as in
the ungauged theory it is sufficient to impose the Bianchi identities
and the four form and dilaton field equations.

\end{document}